\documentclass[reprint,showkeys,showpacs]{revtex4-1}
\usepackage{epsfig}
\usepackage{graphicx}
\usepackage[tbtags]{amsmath}

\begin{document}

\title{Experimental Realization of a Four-Photon Seven-Qubit Graph State for One-Way Quantum Computation}
\date{\today}
\author{Sang Min Lee}
\affiliation{Korea Research Institute of Standards and Science, Daejeon 305-340, Republic of Korea}
\author{Hee Su Park}
\email[Corresponding author. ]{hspark@kriss.re.kr}
\affiliation{Korea Research Institute of Standards and Science, Daejeon 305-340, Republic of Korea}
\author{Jaeyoon Cho}
\altaffiliation[Present address: ]{QOLS, Blackett Laboratory, Imperial College London, London SW7 2BW, UK}
\affiliation{Korea Research Institute of Standards and Science, Daejeon 305-340, Republic of Korea}
\author{Yoonshik Kang}
\affiliation{Korea Research Institute of Standards and Science, Daejeon 305-340, Republic of Korea}
\author{Jae Yong Lee}
\affiliation{Korea Research Institute of Standards and Science, Daejeon 305-340, Republic of Korea}
\author{Heonoh Kim}
\altaffiliation[Present address: ]{Department of Physics, University of Ulsan, Ulsan 680-749, Republic of Korea}
\affiliation{Korea Research Institute of Standards and Science, Daejeon 305-340, Republic of Korea}
\author{Dong-Hoon Lee}
\affiliation{Korea Research Institute of Standards and Science, Daejeon 305-340, Republic of Korea}
\author{Sang-Kyung Choi}
\email[Corresponding author. ]{skchoi@kriss.re.kr}
\affiliation{Korea Research Institute of Standards and Science, Daejeon 305-340, Republic of Korea}

\begin{abstract}
We propose and demonstrate the scaling up of photonic graph state through path qubit fusion. Two path qubits from separate two-photon four-qubit states are fused to generate a two-dimensional seven-qubit graph state composed of polarization and path qubits. Genuine seven-qubit entanglement is verified by evaluating the witness operator. Six qubits from the graph state are used to execute the general two-qubit Deutsch-Jozsa algorithm with a success probability greater than 90$\%$.
\end{abstract}

% 03.67.Bg Entanglement production and manipulation
% 03.67.Lx Quantum computation architectures and implementations
% 42.50.Dv Quantum state engineering and measurements in quantum optics
% 42.50.Ex  Optical implementations of quantum information processing and transfer in quantum optics

\maketitle

Graph states are the essential resource for one-way quantum computation (1WQC)~\cite{Raussendorf, Briegel}. This involves the preparation of qubits entangled in the shape of a graph followed by sequential measurements on its local qubits. As the computational capacity of a graph depends on size as well as structure, much effort has been made to increase the number of qubits in a graph state. A standard approach using photons is to apply entangling gates between basic two-qubit graphs produced by spontaneous parametric down-conversion (SPDC)~\cite{Browne}. For example, 1WQC with six-qubit graph states has been experimentally demonstrated with a combination of three SPDC photon pair sources~\cite{Lu}. Entanglement of more than six photons has not been feasible because of the limited SPDC efficiency.

Schemes to encode more than one qubit per photon have been developed to further increase the number of qubits given the limitation on the number of photons. Single-photon two-qubit (1P2Q) and single-photon three-qubit (1P3Q) schemes using both polarization and photonic path to encode qubits have realized six-qubit graph states~\cite{Gao2, Vallone2}. However, their scalability is limited by the the difficulty of maintaining the long-term phase stability of path qubits. For this reason, the use of path qubits has previously been limited to either two-photon experiments that can be completed within a short time frame~\cite{Kwiat, Vallone3, Chen, Vallone2, Park}, or where path qubits are added as dangling nodes connected to polarization qubits at the final step of generating a graph~\cite{Gao1, Gao2, Kalasuwan}.

Our 1P2Q approach is to design a scheme that implements a fusion of path qubits from different photons with sufficient stability. We demonstrate the realization of a four-photon seven-qubit (4P7Q) graph state by combining two separate two-photon four-qubit (2P4Q) linear graph states through a path qubit fusion gate. Successful fusion of the two 2P4Q states results in genuine seven-qubit entanglement, which can be verified by evaluating the entanglement witness. The fused graph state is sufficiently large with the requisite structure to demonstrate a small-scale quantum algorithm such as the 1WQC protocol for the two-qubit Deutsch-Jozsa algorithm~\cite{Tame2}.

%figures

\begin{figure*}[ht]
\centerline{\includegraphics[scale=0.62]{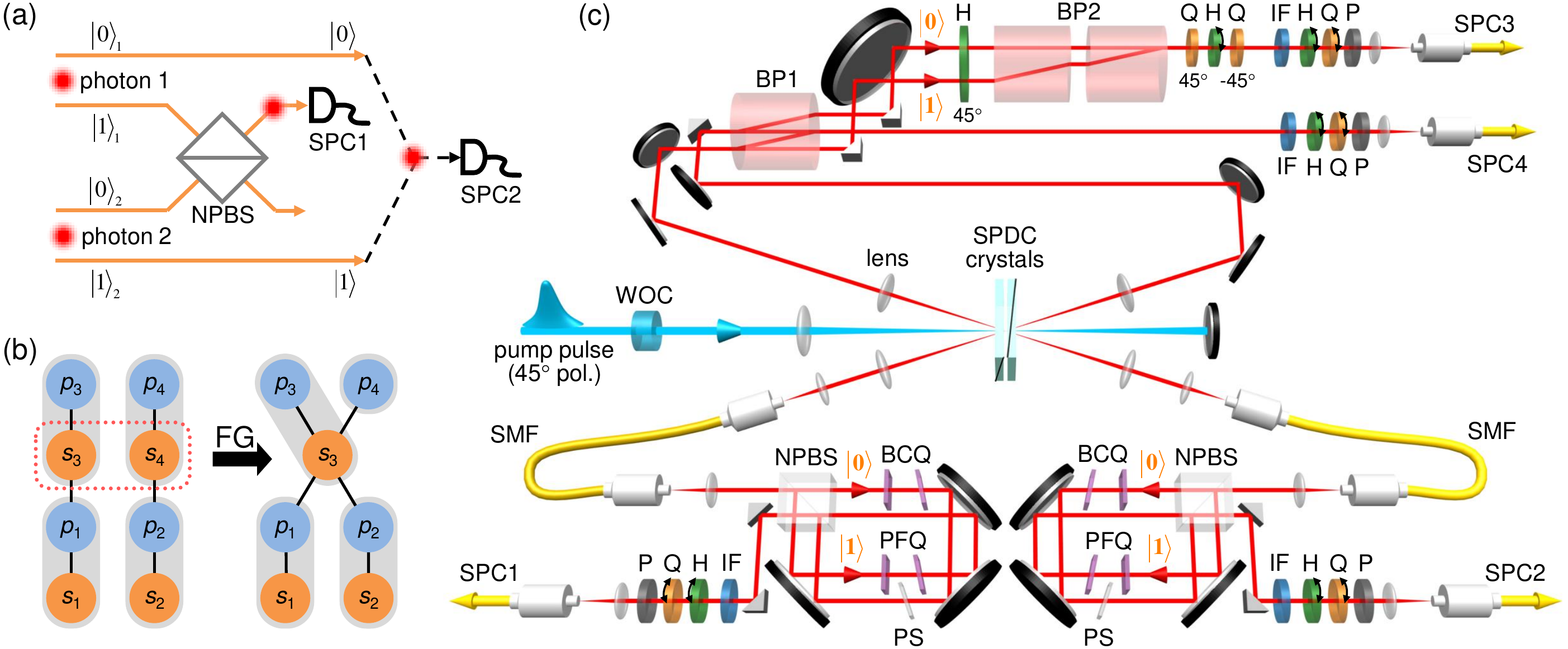}}
\caption{(color online). Seven-qubit graph state generation and measurement. (a) Schematic of path qubit fusion gate. (b) Generation of four-photon seven-qubit graph state from two two-photon four-qubit linear graph states. Polarization qubit $p_i$ and path (spatial) qubit $s_i$ are encoded in photon $i$. FG: fusion gate. (c) Overall experimental setup. BP: birefringent prism, Q: quarter-wave plate, H: half-wave plate, P: polarizer, SMF: single-mode fiber, NPBS: non-polarizing beam splitter, BCQ: birefringence-compensating quartz crystal, PFQ: polarization-flipping quartz crystal, PS: phase shifter, WOC: walkoff compensator, SPC: single-photon counter (PerkinElmer SPCM-AQ4C), IF: interferometer filter (10 nm for SPC1 and SPC2, 5 nm for SPC3 and SPC4). The angles of the wave plates denote the direction of the slow axis with respect to the horizontal.}
\label{f1}
\end{figure*}

Figure 1(a) shows the conceptual scheme of the proposed path qubit fusion gate. The structure in Fig. 1(a) has the same physical topology as the type-I fusion gate proposed for polarization qubits~\cite{Browne}. Two input photons propagate along two paths that represent $|0\rangle_i$ and $|1\rangle_i$, where $i$ denotes the $i$-th photon. The paths $|1\rangle_1$ and $|0\rangle_2$ are superposed by a 50:50 non-polarizing beam splitter (NPBS). Post-selection of the case where the two photons arrive SPC1 and SPC2, respectively, corresponds to the projection $|0\rangle \langle00|_{12} + |1\rangle \langle11|_{12}$ with the initial path $|0\rangle_1$ and $|1\rangle_2$ respectively becoming the paths $|0\rangle$ and $|1\rangle$ of the fused path qubit. When the two photons carry both polarization qubits and path qubits, the two polarization qubits are exchanged depending on whether the fused path qubit is $|0\rangle$ or $|1\rangle$. In other words, the fusion gate shown in Fig. 1(a) consists of a controlled swap operation on the polarization qubits followed by the fusion of the path qubits. Here, the fused path qubit becomes the control qubit. Therefore, to realize a pure fusion operation, the fusion gate should be applied to two photons whose polarization qubits are symmetric and unaffected by the swap operation, as explained below for our scheme.

This fusion gate is applied to path qubits in the middle of two linear 2P4Q states. We briefly describe our 2P4Q state generator. (Details can be found in~\cite{Park}). Initially, two photons are prepared in the Bell state $(|HH\rangle + |VV\rangle)/\sqrt{2}$. The separate paths of the two photons are respectively split by a polarizing beam splitter (PBS) and an NPBS. One output path from the NPBS passes through a half-wave plate (HWP) that interchanges horizontal and vertical polarizations. The linear polarizations along $-45^\circ$ and $45^\circ$ becomes the polarization qubits $|0\rangle$ and $|1\rangle$, respectively; the PBS and the NPBS each have two output paths that respectively represent the logical path qubits of each photon. The fusion gate applied to PBS-side path qubits located in the middle of separate 2P4Q graphs generates a 4P7Q state as depicted in Fig. 1(b).

Figure 1(c) shows the overall experimental setup, from the generation of separate 2P4Q states to the detection of the 4P7Q graph state. The Bell-state photon pairs are generated by type-I SPDC in cascaded BBO crystals~\cite{Kwiat2, Nambu}. The pre-walkoff-compensated~\cite{Nambu} pump laser (wavelength 390 nm, pulse duration 200 fs, repetition 76 MHz, average power 300 mW) propagates in a double-pass geometry through the crystals and generates two photon pairs, in the forward and backward directions, respectively. The four photons are coincidence counted by four single photon counters (SPC1$\sim$SPC4). The photon detected by SPC\hspace{0.05cm}$i$ is henceforth labeled as photon $i$.

High phase stability for the path qubits is achieved by replacing the PBS used in the original 2P4Q state generator~\cite{Park} with a birefringent prism (BP1) that induces polarization-dependent walkoff. BP1 serves a twofold function: (i) generation of two 2P4Q graph states through polarization-dependent beam separation, and (ii) path qubit fusion by combining two separate beam paths for distinct photons. Path combination by BP1 is possible because the respective polarizations in the two paths are mutually orthogonal. The use of a BP instead of an NPBS for path qubit fusion has the benefit of obviating the $50\%$ photon loss associated with the NPBS. After the fusion operation of BP1, the photon sent to SPC4 along the combined path (photon 4) carries only a polarization qubit. The remaining two paths that lead to SPC3 correspond to paths $|0\rangle$ and $|1\rangle$ for photon 3.

The other two photons, photons 1 and 2, are sent along paths directed to detectors SPC1 and SPC2, respectively. Each photon is spatial-mode-filtered by a single-mode fiber (SMF) before entering an NPBS that splits the incoming path into a Sagnac interferometric configuration. One output path from each NPBS passes through a pair of birefringent quartz crystals, BCQs, whose optic axes are horizontally and vertically aligned, respectively, to compensate for unwanted birefringence from optical components~\cite{Park}. Polarization flips, required for 2P4Q state generation, are performed by a pair of birefringent quartz crystals (PFQs) whose optic axes are aligned along 45$^\circ$ or -45$^\circ$ from the horizontal axis. The PFQ crystals are tilted to yield a combined birefringence equivalent to an HWP with slow axis aligned along 45$^\circ$.

The measurement of the polarization qubits $p_1\sim p_4$ follows standard procedure with the use of HWPs, quarter-wave plates (QWPs), and polarizers. Path qubit measurements, however, are configured differently depending on the photon. The path qubits $s_1$ and $s_2$ for photons 1 and 2 are each projected and measured with a Sagnac interferometers whose NPBS initially splits the incoming path and subsequently recombines the interfering paths, as illustrated in the lower part of Fig.~1(c). The path qubit is projected to $|0\rangle$ or $|1\rangle$ by blocking either path with a shutter (not shown in Fig.~1(c)), or projected to the superposed state $(|0\rangle + e^{i \phi}|1\rangle)/\sqrt{2}$ by adjusting the phase $\phi$ with a tiltable 1-mm-thick glass plate (PS).

The path qubit $s_3$ is measured by applying a birefringent prism (BP2) to combine the two paths for photon 3. The path combination is preceded by an interchange of the horizontal and vertical polarizations performed by an HWP before BP2, and followed by another interchange of polarizations through a QWP-HWP-QWP sequence. This double-swap of polarizations has been introduced to match the path lengths for photons 3 and 4: matching the lengths of the two paths for either photon ensures path-length matching for the other photon. Projection to $|0\rangle$ or $|1\rangle$ is done by a shutter located in front of BP2. Rotation of the HWP between the two QWPs changes the phase $\phi$ of the superposition state because the polarization of the $|0\rangle$ ($|1\rangle$) path is fixed as horizontal (vertical). Path combination with a BP rather than an NPBS projects can a photon to a polarization-path entangled state instead of a product state as required for independent measurements of polarization and path qubits. When the polarization state and the path state are respectively projected onto $|p\rangle = a|H\rangle_p + b|V\rangle_p$ and $|s\rangle = c|0\rangle_s + d|1\rangle_s$, where $a$, $b$, $c$, $d$ are constants, the state of photon 3 corresponds to $ac|H\rangle_p |0\rangle_s + bd|V\rangle_p |1\rangle_s$ instead of $|p\rangle |s\rangle$. However, BP1 suppresses $|V\rangle_p |0\rangle_s$ and $|H\rangle_p |1\rangle_s$, hence the projection to $ac|H\rangle_p|0\rangle_s + bd|V\rangle_p|1\rangle_s$ is equivalent to the measurement of $|p\rangle |s\rangle$ within experimental uncertainty.

The operation of the experimental setup is critically dependent on successful path qubit fusion. A stable phase is maintained in our setup by designing the interfering paths to share most of their optical components as shown in Fig. 1(c). The phase stability for path qubits $s_1$ and $s_2$ is maintained by constructing Sagnac interferometers for the paths.

The stability of the fused path qubit is tested by measuring the coherence between the $|0\rangle$ and $|1\rangle$ states of $s_3$. All the polarization qubits adjacent to path qubit $s_3$ are projected to $|0\rangle$, which ideally projects qubit $s_3$ to $|+\rangle$, and the path qubits $s_1$ and $s_2$ are projected to $|+\rangle$ to maximize the four-photon count rate. The coherence of $s_3$ is measured by four-photon coincidence counts with varying relative phase as shown in Fig.~2. The coherence between $|0\rangle$ and $|1\rangle$ is visibly maintained for a total measurement time of 5 hours (23 data points, 800 s each). The non-ideal interference visibility of $0.49 \pm 0.03$ is ascribed to imperfect initial Bell-states (concurrence $\gtrsim 0.9$) and spectral/temporal impurity of photons~\cite{Huang}.

%figures

\begin{figure}[t]
\centerline{\includegraphics[scale=0.7]{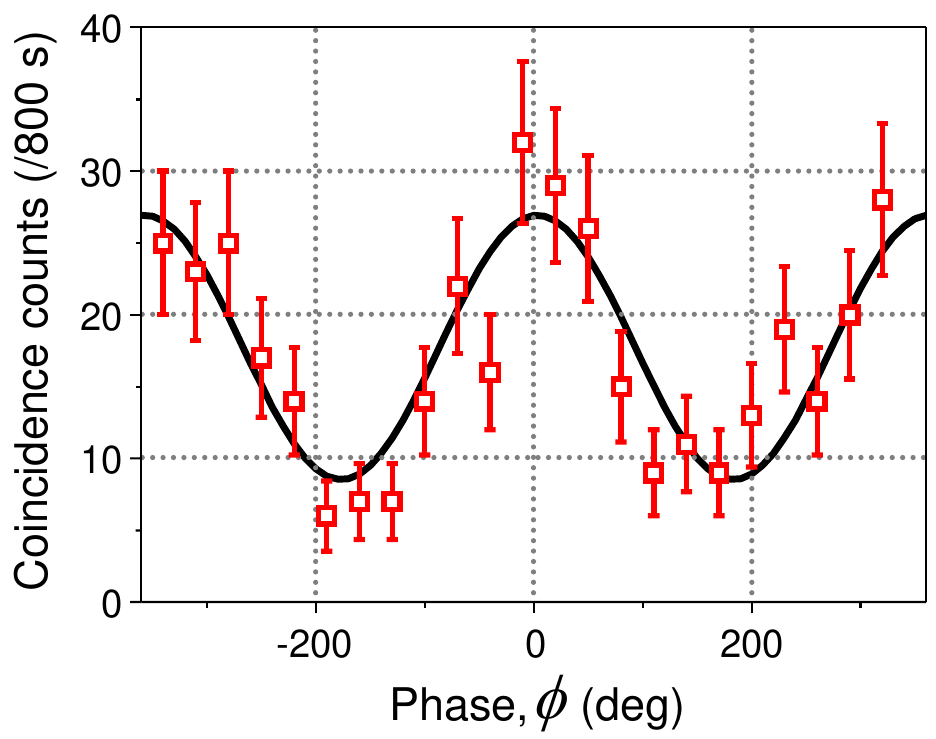}}
\caption{Interference between the $|0\rangle$- and $|1\rangle$-components of fused path qubit. Fourfold coincidence counts in 800 s . Qubits $p_1 \sim p_4$ and $s_1 \sim s_2$ projected to $|0\rangle$ and $|+\rangle$, respectively; Qubit $s_3$ projected to $(|0\rangle + e^{i \phi} |1\rangle)/\sqrt{2}$. Error bars denote $\pm\sqrt{\text{counts}}$.}
\label{f2}
\end{figure}

A definite test of the 4P7Q state is the measurement of the entanglement witness defined as~\cite{Toth2}
\begin{gather}
W=3I-2 \left[ \prod_{i=1}^{3} \frac{S(s_i)+I}{2} + \prod_{i=1}^{4} \frac{S(p_i)+I}{2} \right] .
\end{gather}
Here, $I$ is the identity matrix, $S(q_i)$ is the stabilizing operator $S(q_i)=X_i\prod_{j }Z_j$ , where the product is over the qubits $\{q_j\}$ adjacent to qubit $q_i$, and $X_i$ and $Z_j$ are Pauli operators. Since our graph state is two-colorable~\cite{Toth}, two measurement configurations ($X$- and $Z$-measurements for polarization and path qubits, respectively, and vice versa) are sufficient to estimate all the stabilizing codes. The raw measurement data are listed in the Supplementary Material~\cite{sup}.  Table 1 shows the measurement results for $S(q_i)$ and $W$. The expectation value of $W=-0.281 \pm 0.069$ is below zero (the separability bound), hence indicates genuine multipartite entanglement for the generated 4P7Q state. The fidelity with the ideal state is estimated to be $F\geq64\%$ from the witness value~\cite{Toth}.

% table 1

\begin{table}[h]
\begin{tabular}{c c c c | c c c c}
\hline \hline
operator & & value & & & operator & & value \\
\hline
$S(p_1)$ & & $1.000\pm 0.0 $ & & & $S(s_1)$ & & $0.925\pm0.022$ \\
$S(p_2)$ & & $0.948\pm0.025$ & & & $S(s_2)$ & & $0.933\pm0.007$ \\
$S(p_3)$ & & $0.974\pm0.018$ & & & $S(s_3)$ & & $0.454\pm0.052$ \\
$S(p_4)$ & & $0.961\pm0.022$ & & & $W$ & & $-0.281 \pm 0.069$ \\
\hline \hline
\end{tabular}
\caption{Expectation values of stabilizing operators and entanglement witness.}
\label{t1}
\end{table}

Our genuinely entangled 4P7Q state is applied to a demonstration of a 1WQC algorithm, specifically the two-qubit Deutsch-Josza algorithm (DJA) developed for an E-shaped six-qubit graph state~\cite{Tame2}. This six-qubit state is prepared by detaching qubit $p_4$ from the seven-qubit graph by projection to $|0\rangle$ as shown in Fig.~3(a). The DJA scheme is implemented according to the procedure shown in Table 2 of \cite{Tame2}. Qubits $p_1$, $p_2$, and $p_3$ constitute an oracle preparing a function and an ancilla, while qubits $s_1$, $s_2$, and $s_3$ reveal the computation result. The DJA is executed for four functions, $f(\{0,1,2,3\}) = \{0,0,0,0\}$, $\{0,0,1,1\}$, $\{0,1,0,1\}$, $\{0,1,1,0\}$, which are labeled (i), (iii), (v), (vii), respectively; the other four functions, $f(\{0,1,2,3\}) = \{1,1,1,1\}$, $\{1,1,0,0\}$, $\{1,0,1,0\}$, $\{1,0,0,1\}$ require exactly the same measurement bases as (i), (iii), (v), (vii), respectively, hence can be omitted without loss of generality.  Depending on which function is selected, a set of measurements ($Y$ or $Z$, $X$, $X$ or $Y$, $Z$) is performed on qubits $p_1$ and $p_2$, $p_3$, $s_1$ and $s_2$, $s_3$, respectively, and feedforward is applied afterwards. The raw data with the measurement bases are listed in the Supplementary Material and the output probability results are shown in Fig.~3(b). The output probability is greater than 90\% when compared with the ideal case where $|s_3\rangle$ is always $|-\rangle$ and $|s_1\rangle |s_2\rangle$ is $|0\rangle |0\rangle$ only for a constant function.

%figures

\begin{figure}[t]
\centerline{\includegraphics[scale=0.215]{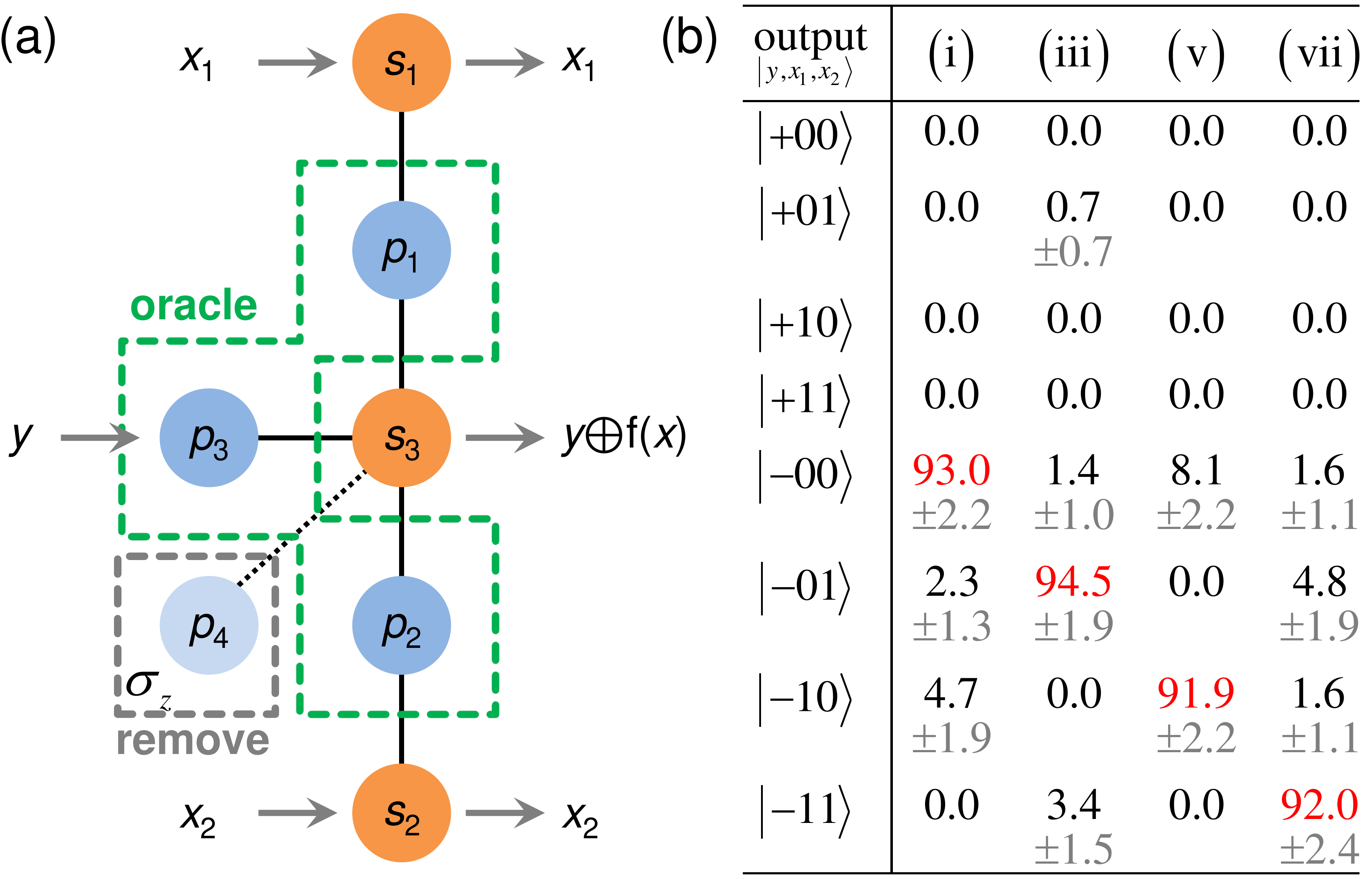}}
\caption{(color online). General two-qubit Deutsch-Jozsa algorithm on the 4P7Q state. (a) Structure of graph state and logic flow. (b) Measured output probability $(\%)$ for ancilla qubit ($y$) and query qubit ($x_1$ and $x_2$). (i), (iii), (v), (vii) are results for functions $f(\{0,1,2,3\}) = \{0,0,0,0\}$, $\{0,0,1,1\}$, $\{0,1,0,1\}$, $\{0,1,1,0\}$, respectively. }
\label{f3}
\end{figure}

We have demonstrated that fusion of path qubits from distinct photons is a feasible approach to generating larger and more complex graph states. The realization of one-way quantum computation (1WQC) with a seven-qubit graph state is enabled by path qubit fusion. Our results extend the previous work on the two-qubit Deutsch-Jozsa algorithm (DJA) executed for two functions~\cite{Vallone1} to encompass all functions that are constant or balanced. To our knowledge, this is the first 1WQC demonstration of the general two-qubit DJA. Our method of combining path qubit fusion with two-photon four-qubit state generation is applicable to other schemes that use path qubits~\cite{Gao1, Kalasuwan}. We expect this strategy for scaling up graph states to be useful for the generation of graph states with structures suitable for other 1WQC algorithms.

\begin{acknowledgments}
This work has been supported by the KRISS project `Single-Quantum-Based Metrology in Nanoscale.'
\end{acknowledgments}

\end{document}